\documentclass[superscriptaddress,preprintnumbers,keywords]{revtex4}
\usepackage{epsfig,dcolumn}
\usepackage{graphicx}
\usepackage{color}

\usepackage{amsmath}
\usepackage{amsfonts}
\usepackage{amssymb}
\usepackage{graphicx}

\newcommand{\be}{\begin{equation}}
\newcommand{\ee}{\end{equation}}

\newcommand{\eea}{\end{eqnarray}}

\def\lsim{\mathrel{\rlap{\lower4pt\hbox{\hskip1pt$\sim$}}
    \raise1pt\hbox{$<$}}}
\def\gsim{\mathrel{\rlap{\lower4pt\hbox{\hskip1pt$\sim$}}
    \raise1pt\hbox{$>$}}}

\usepackage{bm} 

\begin{document}

\title{Triangle Singularities and XYZ Quarkonium Peaks} 

\author{Adam P. Szczepaniak} 
 \affiliation{ 
  Department of Physics, Indiana University, Bloomington, IN 47405, USA } 
\affiliation{ Theory Center, Thomas Jefferson National Accelerator Facility, \\
12000 Jefferson Avenue, Newport News, Virginia 23606, USA} 
  \affiliation{ 
 Center for Exploration of Energy and Matter, Indiana University, Bloomington, IN 47403, USA}

\preprint{JLAB-THY-15-1999}


\begin{abstract}
We discuss analytical properties of partial waves derived from projection of a 4-legged amplitude with crossed-channel exchanges  in the kinematic region of the direct channel that corresponds to the XYZ peaks in charmonium and bottomonium.  We show that in general partial waves can develop anomalous branch points in the vicinity of the direct channel physical region. 
We show that this effect only occurs  if masses of resonances in the crossed channel are in a specific, narrow range.  We estimate the size of threshold enhancements originating from these anomalous 
 singularities in reactions where the  $Z_c(3900)$ and the $Z_b(10610)$ peaks have been observed. 
  \end{abstract}

\maketitle

\section{Introduction}
\label{sec:introduction}
 
There is significant interest in the physics of heavy quarkonia stimulated by discoveries of 
  narrow peaks in the spectrum. Such peaks may indicate existence of new hadrons. A recent review of the experimental situation and of the various theoretical models can be found, for example, in ~\cite{Esposito:2014rxa}.  While the quark model provides a remarkably accurate  description of the heavy quarkonium spectrum, these new peaks, also referred to as the XYZ states, appear at masses that do not, in a natural way, derive from the quark model~\cite{Swanson:2006st}.  Several of these peaks have been observed in invariant mass distributions of meson pairs that contain one heavy quarkonium, {\it e.g.}  the $J/\psi$ or the $\Upsilon$,  and one light meson. The possibility that these new hadrons are therefore multi-quark bound states has been explored, for example in \cite{Maiani:2014aja,Faccini:2013lda}.  The peaks appear near thresholds for production of meson pairs with open flavor, {\it e.g.} $D\bar D^*$ or $B\bar B^*$, and for this reason it has also been suggested that binding between the two flavored mesons may be responsible for generating some of the XYZ's~\cite{Tornqvist:2004qy,Stapleton:2009ey,Wang:2013cya,He:2013nwa}. It should be recognized, however, that distinction between multi-quark and hadron bound states is complicated~\cite{Jaffe:2007id} and in any case requires sophisticated amplitude analysis and  precise data  \cite{Morgan:1993td}. 
 It has also been suggested ~\cite{Bugg:2011jr} and studied in ~\cite{Chen:2011pv,Chen:2011xk,Chen:2013coa,QWang,Guo:2014qra,Swanson:2014tra,Pak}  that coupling to nearby open channels may produce peaks even without presence of new hadrons. In the mathematical language of amplitude analysis the analogous statement would be that threshold cusp may be enhanced not only by a nearby pole but by a carefully arranged set of branch points, also referred to as anomalous thresholds. Amplitude poles correspond to physical particles (stable or unstable) while cuts represent the effect of "forces", {\it i.e.} exchanges of known particles in the crossed-channels. If the XYZ peaks were due to the latter it could potentially weaken the case for new hadrons. To address this issue it is therefore important to perform a systematic analysis of the "forces" and the associated singularities.

\section{The Model} 
\label{sec:model} 
     We consider a decay of a heavy object of mass $M$, hereafter referred to as the $M$ particle, to a quasi stable heavy state of mass $M'$  and a pair of degenerate, light mesons of mass 
    $\mu$. We are interested in the energy dependence of the low spin partial waves,  in particular the $S$-waves, in the $M'\mu$ channel.  In this section we discuss a generic case and in the section that follows 
 we consider threshold behavior in $J/\pi \pi$ and $\Upsilon(1S)\pi$ production from  
     $Y(4260) \to J/\psi \pi\pi$ ~\cite{Ablikim:2013mio,Liu:2013dau,Xiao:2013iha} and $\Upsilon(5S) \to \Upsilon(1S) \pi\pi$ \cite{Collaboration:2011gja,Garmash:2014dhx} decays, respectively. 
   
  The reaction of interest is $M(p_1) \to M'(p_3) + \mu(p_2) + \mu(p_4)$ with $p_i$ referring to the 4-vectors. Ignoring spin, the reaction amplitude,  $A(s,t)$ is a scalar function of  two independent Lorentz invariant Mandelstam variables, which we choose as,  $s=(p_1 - p_2)^2$ and $t=(p_1 - p_3)^2$. They correspond to the invariant mass squared of the $M'(p_3) \mu(p_4)$ pair and momentum transfer squared between the two heavy mesons, respectively. The amplitude in the kinematics of the decay region, $s>(M'+ \mu)^2$, $t>4\mu^2$ can be obtained by analytical continuation in $t$  of the  amplitude describing the $s$-channel scattering process,  $M(p_1) + \mu(-p_2) \to M'(p_3) + \mu(p_4)$.  It is the $t$-channel singularities of the latter that are responsible for anomalous singularities of the $s$-channel partial waves. Similar analysis applies to the $u$-channel where $u = (p_1 - p_4)^2$. Bose symmetry requires the amplitude to be  $s  \leftrightarrow u$ symmetric. 

   The $s$-channel partial waves, $A_l(s)$, describe production of the $M'\mu$ system in a state of fixed angular 
   momentum, $l$. Ignoring spin of the external particles, partial waves are the coefficients in expansion 
    of $A(s,t)$ in a series of Legendre polynomials, 
    \begin{equation} 
   A(s,t) = 16\pi \sum_{l=0}^\infty (2l+1) A_l(s) P_l(z_s).  
   \end{equation} 
   The argument of the rotational functions, $z_s = z_s(s,t)$, is the cosine of the scattering angle in the center of mass frame of the $s$-channel.  
 When considered as complex functions of the invariant mass squared, $s$, the partial waves $A_l(s)$ have branch points at all $s$-channel production thresholds.  The cuts associated with these branch points are located on the positive real $s$ axis and unitarity determines the discontinuity across the cuts. These cuts are referred to as the right hand cuts. 
  Through the partial wave projection, 
   \begin{equation} 
   A_l(s) = \frac{1}{ 32 \pi} \int_{-1}^1 dz_s P_l(z_s) A(s,t), 
   \end{equation} 
   threshold singularities of $A(s,t)$ in the $t$ and the $u$ channel also lead to branch points in the complex $s$ plane.  The  associated cuts are generically referred to as the left hand cuts, even though, as it will be the case here, they 
    don't always lie on the negative real $s$-axis.  In the following we consider only the $S$-waves ($l=0$) and drop the angular momentum subscript on the partial waves.  Higher partial waves are suppressed at thresholds due to the  angular momentum barrier factors.    
    
     In the cases considered here,  $\sqrt{s}$ is of the order of several GeV and there are several open 
      channels  contributing  to the right hand side discontinuity. As long as their thresholds are far from the region of interest they will not lead to rapid variations in $s$. It is therefore sufficient to focus on the channels, which open near the XYZ production thresholds. 
    For simplicity we assume that the relevant channel consists of two identical mesons with mass $m$. In the cases considered in the following section these will correspond to the $B\bar B^*$ or the $D \bar D^*$ state. Here, for simplicity,  
     we ignore the mass difference between the pseudoscalar and vector mesons. We assume that  masses satisfy the following relation, $M \gsim 2m$ and $2m > M'$, {\it i.e.}  we take mass of the decaying/produced quarkonium   to be slightly above/below the open flavor threshold.   The mass of the light particle, $\mu$ is much smaller compared to masses of the other mesons that all contain heavy quarks, $\mu << m,M,M'$, so in the following we take $\mu=0$. For $\mu\ne 0$  the amplitude has a slightly different analytical structure. The difference, however, does not affect the behavior near the $2m$  threshold.  
    
      The contribution of the $2m$ channel to the discontinuity of  the $M + \mu \to M' + \mu$ $S$-wave amplitude, 
       is given by 

\begin{equation} 
\Delta A(s) = B^*(s) \rho(s) C(s) \theta( s- 4m^2).  \label{dis} 
\end{equation} 
Here $B(s)$ and $C(s)$ are the $S$-wave projections of the amplitudes $B(s,t(s,z_s))$ and  $C(s,t(s,z_s))$  describing the 
 reactions, $M'(p_3) + \mu(p_4) \to m(q_1) + m(q_2)$ and  $M(p_1) + \mu(-p_2) \to m(q_1) + m(q_2)$, respectively. The phase space factor $\rho(s)$ is given by $\rho(s) = \sqrt{1 - 4m^2/s}$. It should be noted that amplitude discontinuity,  $\Delta A(s) = (A(s+ i\epsilon) - A(s - i\epsilon))/2i$ is in general a complex function of $s$ and not equal to $\mbox{Im}A(s)$, 
  as sometimes assumed~\cite{Bugg:2011jr,Swanson:2014tra}. This has important consequences when considering  phase of the 
   amplitude, which is often used to discriminate between resonant and non-resonant behaviors.

  We reconstruct $A(s)$ from its discontinuity approximated by Eq.~(\ref{dis}) since, in the region of interest, the difference between $\Delta A(s)$ and the true discontinuity is expected to be a smooth function of $s$, 
\begin{equation} 
A(s) = \frac{1}{\pi} \int_{s_{tr}} ds' \frac{B^*(s') \rho(s') C(s')}{s' -s}. \label{unit} 
\end{equation} 
where $s_{tr} = 4m^2$.  If $A(s)$ is to have a strong dependence on $s$, the numerator under the integral has to vary rapidly with $s$.  Rapid variations are determined by nearby singularities.  As discussed above, if direct channel bound states are excluded, the only physically allowed singularities of partial waves are branch points. Thus  in order for the integral in Eq.~(\ref{unit}) to develop a singularity the branch points of the numerator have to either appear at the endpoint of the integration region or pinch the integration contour. 
 
 We note here, that in \cite{Swanson:2014tra} the amplitudes $B(s)$ and $C(s)$ were  chosen proportional to an exponential, $\exp(-(s-s_{tr})/s_0)$ with $s_0 = O(1\mbox{GeV}^2)$. Such behavior was motivated by a quark model. Non-relativistic quark model calculations often involve gaussian wave functions and one could imagine that a calculation, in which the scattering amplitude is derived from a diagram involving quark exchanges between interacting mesons, would produce energy dependence that falls off exponentially. Such a model, at best,  is valid for non-relativistic relative momenta and has incorrect crossing-properties that prevents it from being used in a dispersive analysis. A dispersive analysis requires that the amplitudes vanish  at infinity in all directions in the complex $s$-plane or, in the worst case, grow polynomially. Instead, the amplitudes in \cite{Swanson:2014tra}  have an essential singularity at infinity.  Such a singularity is unphysical, it violates causality, which 
  requires the amplitudes to be  polynomially bound.  If a proper Lorentz covariant quark model was constructed the quark exchange mechanism would correspond  to $u$-channel exchanges that do not overlap with exchanges of normal, {\it i.e.} quark-antiquark, mesons. Such "forces" lead  to $s$-channel partial wave amplitude singularities located far to the left from $s$-channel unitary cuts and as such do  not produce enhancements in the $s$-channel physical region. For example  the form  $\exp(-(s-s_{tr})/s_0)$ might be  replaced by $(s_{tr} + s_0)/(s + s_0)$.  The latter vanishes in all directions at infinity and with  $s_0 > 0 $ it  is often used to approximate singularities located far away to the left 
   from the physical region. When such a form factor is used in  Eq.~(\ref{unit}) (in place of the $B$ and $C$) it  results in a normal, not enhanced  threshold cusp.

In the following we show that $A(s)$ develops a  pinch singularity through  $C(s)$ but not for through  $B(s)$. Since $C(s)$ is the $s$ channel, S-wave projection of the $M + \mu \to m + m$ amplitude 
\begin{equation} 
C(s) = \frac{1}{ 32 \pi} \int_{-1}^{1} dz_s C(s,t(s,z_s))
\end{equation} 
we use the $t$-channel dispersion relation to write, 
\begin{equation} 
C(s) =  \frac{1}{32\pi^2 } \int_{\lambda^2_{min}}^\infty d\lambda'^2  C_1(s,\lambda'^2) Q(s,\lambda'^2) 
\end{equation} 
where 
\begin{equation} 
Q(s,\lambda^2) \equiv  \int_{-1}^1 \frac{dz_s }{\lambda^2 - t(s,z_s)}. \label{t} 
\end{equation} 
Here $C_1$ is the $t$-channel discontinuity of $C(s,t)$ at fixed-$s$ and $\lambda^2_{min}$ is the location of the lowest mass singularity. 
The lowest mass that can be exchanged in the $t$-channel is the pseudoscalar or vector (here, for simplicity, assumed degenerate and stable) open flavor, {\it e.g.} $D$ or $D^*$, meson.  In this case $C_1(s,\lambda^2) = G \delta(\lambda^2 - m^2)$ (ignoring spin of the exchanged particle) and the constant  $G$ is given by the product of couplings of the exchanged meson to the $(M m)$ and $(m \mu)$ meson pairs. For larger $\lambda^2$ unitarity in the $t$-channel, relates the discontinuity $C_1(s,\lambda^2)$ to the product of amplitudes $M + m  \to  m + (n \mu)$ and $ m + \mu \to m + (n \mu)$ representing the interaction between the external meson pairs $(M m)$ and $(m \mu)$,  and intermediate states containing,  besides the heavy meson, other hadrons, {\it e.g.} $n$ light mesons, $(n\mu)$.  It will be shown below that the relevant singularities of $C(s)$  originate from a small interval in $\lambda^2$ that corresponds to $t$-channel intermediate states containing a few particles.  Complicated $t$-channel intermediate states are therefore not relevant when looking for sources of  threshold cusp enhancement. 

 The key is the function $Q(s)$ defined in Eq.~(\ref{t}). It is the $S$-wave projection of the $t$-channel amplitude describing an exchange of an object of mass $\lambda$.  As a function of $s$ and $z_s$ the momentum transfer $t$ is given by 
\begin{equation} 
t(s,z_s) = M^2 + m^2 - \frac{s + M^2}{2} + \frac{(s-M^2)\sqrt{s - 4m^2}}{2\sqrt{s}} z_s. 
\end{equation} 
   Typically, the $s$-channel partial wave projection of a 
  $t$-channel exchange leads to an amplitude with branch points located to the left from the $s$-channel thresholds and therefore outside the $s$-channel physical region. 
Singularities in $Q(s)$ near the $s$-channel physical region can appear, however, because of unequal mass kinematics. 
  Formally $Q(s)$ is given  in terms of the $l=0$ Legendre function of the second type, 
 \begin{equation}
Q(s,\lambda^2) =  \frac{2\sqrt{s}}{(M^2-s)\sqrt{4m^2-s}} \log\frac{\lambda^2 - t_-(s)}{\lambda^2 - t_+(s)}  \label{C} 
\end{equation}
where $t_\pm(s) \equiv t(s,\pm 1)$. Determination of $Q(s)$ on the physical sheet  where the dispersion relation, 
Eq.~(\ref{unit}),  is evaluated, corresponds to an appropriate choice of branch cuts of the logarithm and the square roots.

\begin{figure}
\centering
\rotatebox{0}{\scalebox{0.25}[0.25]{\includegraphics{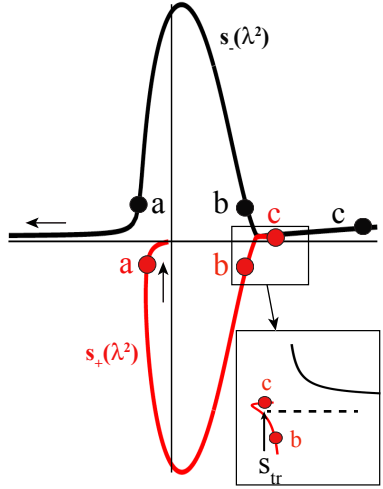}}}
\caption{ Motion of the $s_\pm(\lambda^2)$ branch points in the complex $s$-plane as a function of $\lambda^2$.  The two arrows indicate the limits as $\lambda^2 \to \infty$.  As $\lambda^2$ decreases 
 $s_-$ moves to the right above the real axis. For $\infty > \lambda^2 > \lambda_b^2$,  $s_+$ moves to the right below the real axis.  For  $\lambda^2_b >  \lambda^2 > \lambda_c^2$, $s_+$ moves to the left below the real axis. As $\lambda^2$ decreases below $\lambda_c^2$, $s_+$  circles the real axis below threshold and moves to the right above the real axis.  
The point  {\it a}  represents is located  near $\lambda_a^2$. Point {\it b} is near and above $\lambda_b^2$. When $M^2$ approaches the real axis, $s_\pm(\lambda_b^2)$ pinch the real axis at  $s= s_b = M^2 m/(M-m)$. 
The point {\it c} is near and below $\lambda^2_c$. For $\lambda^2_c >  \lambda^2$ branch points $s_\pm(\lambda^2)$ are located on the same side of the real axis and do not produce a singularity in $A(s)$ when $s$ approaches real axis from above. Pinching occurs for $\lambda^2$  in region {\it iii)}, {\it i.e.}
  $\lambda^2_b >  \lambda^2 > \lambda_c^2$.   } 
\label{spm}
\end{figure}

\begin{figure}
\rotatebox{0}{\scalebox{0.25}[0.25]{\includegraphics{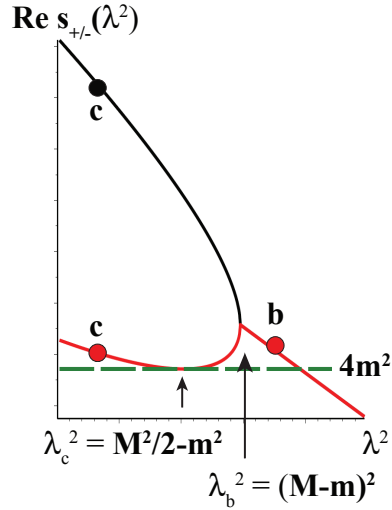}}}
\caption{Behavior of the real parts of $s_\pm(\lambda^2)$ as a function of $\lambda^2$ in the region corresponding to the
 box shown in Fig.~\ref{spm}.  } 
\label{respm}
\end{figure}

 For $\lambda^2$ real, the function $Q(s)$ on the physical sheet  is obtained by continuing in $M^2$ from above the real axis, $M^2 \to M^2 + i\epsilon$~\cite{mandelstam}. The function has four branch points, $s=0,-\infty$ and $s_\pm(\lambda^2)$, given by   the two independent solutions of the equation  $t_\pm(s_\pm(\lambda^2))  = \lambda^2$. The motion of the $s_\pm$ branch points as a function of $\lambda^2$ is shown in Fig.~\ref{spm} and Fig.~\ref{respm}. 
 There are four  characteristic regions, which we enumerate here, 
 
 \begin{enumerate} 
\item[{\it i)}]  $\lambda^2 >  \lambda_a^2 = (M + m)^2$, 
\item[{\it ii)}]  $\lambda_a^2 > \lambda^2 >  \lambda_b^2 = (M - m)^2$, 
\item[{\it iii)}] $\lambda^2_b > \lambda^2 > \lambda_c^2 = M^2/2 - m^2$,
\item[ {\it iv)}] $\lambda^2_c > \lambda^2$.
\end{enumerate} 

 For large values of $\lambda^2$, {\it i.e} in region {\it i)}, Eq.~(\ref{C}) represent a genuine "force" effect of a particle exchange. Branch points of $Q(s)$ are  located on the negative $s$-axis with cuts extending between the point $s= s_+(\lambda^2)$ and $s=0$ and the points $s = -\infty$ and $s= s_-(\lambda^2)$.  As illustrated by the arrows in Fig.~\ref{spm}, when $\lambda^2 \to \infty$,  $s_+(\infty)  =  0$ and $s_-(\infty) = -\infty$ {\it i.e.}, cuts disappear and the exchange reduces to a point-like interaction.  
 At $\lambda^2 = \lambda^2_a$, the two $s_{\pm}$ branch points collide and as  $\lambda^2$ decreases into region {\it ii)} 
  the branch points move into the complex plane. In this region the left hand cut extends over the entire negative real axis and along an arc in the complex plane between $s_+$ and $s_-$. As $\lambda^2$ approaches $\lambda_b^2$ from above, $s_{+/-}$ approach the real axis from below and above, respectively. This happens in the physical region of the $s$-channel.   As long as $M^2$ is slightly above the real axis the two branch points $s_\pm$ never cross the real $s$-axis. 
  In the limit  $\epsilon \to 0$ they pinch it instead. This is the origin of a singularity of $A(s)$. As $\lambda^2$ decreases towards $\lambda^2_c$,  $s_+$ moves to the left, circles the threshold at $\lambda^2 = \lambda^2_c$ and, avoiding the unitary cut,  when $\lambda^2$ decreases below $\lambda^2_c$ into  region {\it iv)},  moves above the real axis. 
      In region {\it iv)} the branch point $s_-$ stays above the real axis and moves to the right as $\lambda^2$  decreases.   For $s$ in the physical region, {\it i.e.} approaching the real axis from above we can draw the following conclusions. 
    Since the $s_\pm$ branch points never cross the unitary cut, the integral in Eq.~(\ref{unit}) is well defined. When  $\lambda^2$ is in region {\it  i)} or {\it ii)} the numerator in Eq.~(\ref{unit}) has singularities far away from the physical region and the only singularity of $A(s)$ is the threshold branch point. For $\lambda^2$ in region {\it iii)} $s$ and $s_+$ are on the opposite sides of the integration contour and  $A(s)$ develops a pinch singularity of a log-type at $s=s_+$.  As $\lambda^2$ decreases below $\lambda^2_c$  the threshold point,  $s=4m^2$ is once more the only singularity of $A(s)$.  
 The real axis singularity for $\lambda^2$ in region {\it iii)} can also be traced to the complex singularities of $A(s)$ for other values of $\lambda^2$. For example, in region {\it ii)} 
 $A(s)$ has a singularity on the second Riemann sheet reached by going down the unitary cut from above. As $s_+$ approaches the real axis from below
  {\it i.e.} $\lambda^2$ moves from region {\it ii)} to {\it iii)}, this second sheet singularity moves towards 
   the real axis.


   \begin{figure} 
   \centering
\rotatebox{0}{\scalebox{0.25}[0.25]{\includegraphics{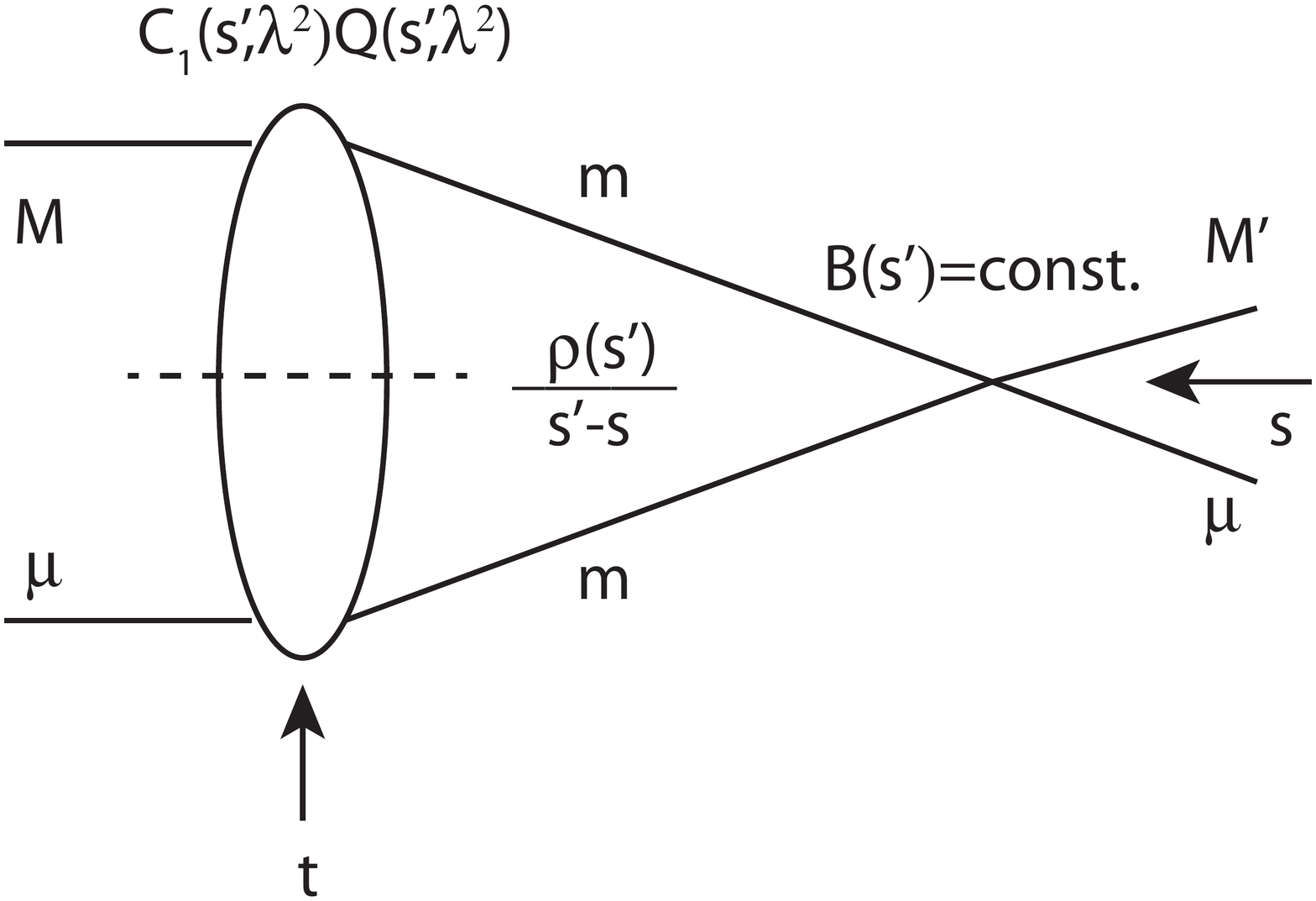}}}
 \rotatebox{0}{\scalebox{0.18}[0.18]{\includegraphics{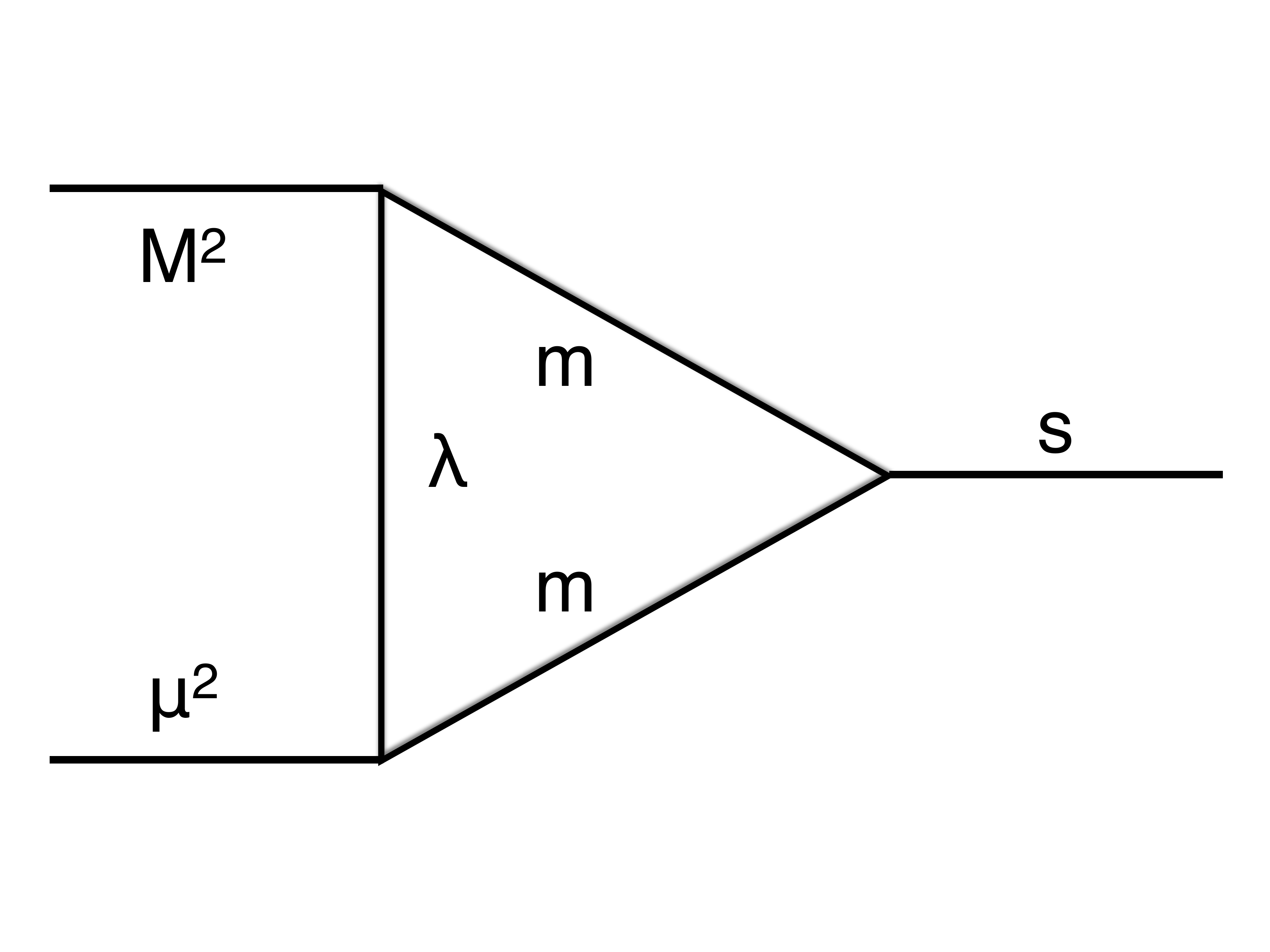}}}
\caption{ Left panel: representation of $A(s)$ given by Eq.~(\ref{example}). Right panel: for real $s$, $A(s)$ is  equivalent (up to arbitrary normalization), to a Feynman triangle diagram in a  scalar theory \cite{Bronzan:1963xn}, $A(s) = \int_0^1 d\alpha_1 d\alpha_2  d\alpha_3 \delta(1 - \alpha_1 - \alpha_2 - \alpha_3) 
 \left[  \alpha_1 \lambda^2  + (\alpha_2 +\alpha_3) m^2 - \alpha_1 \alpha_3 
 \mu^2  - \alpha_2 \alpha_3 s  -\alpha_1 \alpha_2 M^2 - i\epsilon\right]^{-1}$. The internal lines represent scalar particles with masses $m$ and $\lambda$ and squares of momenta of the three  external lines are $M^2$, $mu^2$ and $s$, respectively. \label{tr} }
\end{figure}

  When $M<2m$, which is the case of amplitude $B(s)$, there is no pinching. This is because the condition 
   $(M - m)^2 > \lambda^2 > M^2/2 - m^2$ requires that the mass of the $t$-channel exchange is smaller than the mass of the lightest open flavor meson, $\lambda < m$. Thus $B(s)$ is expected to be a smooth function and in the following we approximate it by a constant. In this approximation,  $A(s)$ given by 
   
   \begin{equation} 
   A(s) = A_0 \int_{4m^2}^\infty  ds' \int_{\lambda_{min}}^\infty d\lambda'^2 \frac{C_1(s',\lambda'^2) \rho(s') Q(s',\lambda'^2) }{s' -s}, \label{example} 
   \end{equation} 
  This  is equivalent to a triangle diagram shown in Fig.~\ref{tr} thus Eq.~(\ref{example})
 has the same singularity structure as the triangle diagram in perturbation theory with the branch points $s_\pm$ corresponding to the leading Landau singularities ~\cite{IA,CS,Bronzan:1963xn}.

\begin{figure}
\hspace*{-.5cm}\rotatebox{270}{\scalebox{0.3}[0.3]{\includegraphics{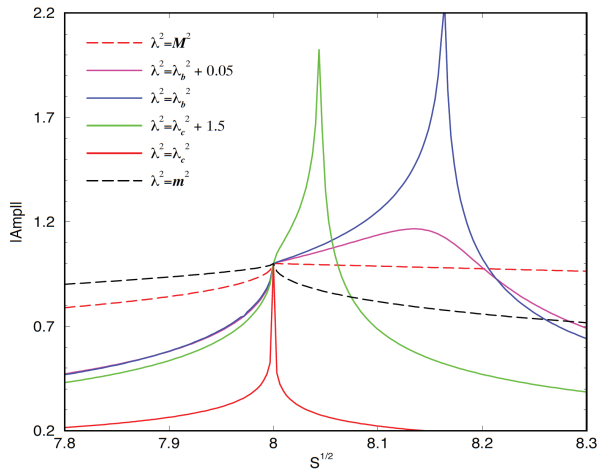}}}
\caption{ Example of $|A(s)|$ obtained using Eq.~(\ref{example}). See the text for interpretation of the results. Color online.} 
\label{l}
\end{figure}

   To illustrate the points discussed above we consider a model with an exchange of a single particle with mass $\lambda^2$. In this case $C_1(s,\lambda'^2) = \delta(\lambda'^2 -\lambda^2)$ and, in arbitrary units, we choose  $M =10$ and $m=4$. In Fig.~\ref{l} we show modulus of $A(s)$ as a function of $s$ for different values of $\lambda^2$. The constant  $A_0$ in 
   Eq.~(\ref{example}) is chosen such that $|A|$ is normalized to unity at threshold, $|A(4m^2)| = 1$. 
For the choice of masses given above, $\lambda_c^2 = 34$ and $\lambda_b^2 = 36$. For $\lambda^2 = 100$, {\it i.e.} in region ${\it ii)}$ the numerator in Eq.~(\ref{example}) has branch points in the $s$-plane far to the left from the threshold and the only $s$-plane singularity of $A(s)$ is the threshold branch point. As $\lambda^2$ decreases and reaches a point slightly above $\lambda_b^2$ the branch points $s_\pm(\lambda)$ approach the real axis. This produces a bump in $s$ close to $s_b   =  M^2 m/(M-m) = 8.165^2$, which originates from the second sheet singularity discussed above. At $\lambda^2=\lambda_b^2$ pinching begins {\it i.e.} in addition to the threshold singularity the amplitude develops a singularity at $s_p=s_b$. As $\lambda^2$ decreases towards $\lambda^2_c$ the $s_+(\lambda)$ branch point moves to the left bringing the pinch singularity closer to threshold. 
When $\lambda^2$ reaches $\lambda^2_c$, the singularity at $s_p$ collides with the threshold singularity  producing a sharp peak at $s=4m^2$. When $\lambda^2$ decreases further the pinch singularity moves away from the real axis to the second Riemann sheet 
reached by moving through the unitary cut from below. This singularity is far away from the physical region leaving the point $s=4m^2$ as the only visible singularity of $A(s)$ in the physical region~\cite{goebel}.  Enhancement of the normal threshold cusp and/or appearance of the pinch singularity near threshold can happen for $\lambda^2$ in a limited range. This is important for phenomenology, since it implies that only a small part of the $t$-channel spectral function is relevant. Physical interpretation of the kinematics corresponding to region  {\it iii)} was given in ~\cite{vv}. 
The strongest effect occurs in the  case $M=2m$ when $\lambda^2_b = \lambda_c^2$ In this case pinch and end-point (threshold) singularities overlap producing effectively a pole of the numerator in Eq.~(\ref{example}) at $s=4m^2$. 
 The numerator behaves like a pole only for $s>4m^2$. To the left of $s=4m^2$ there is a branch 
 cut of infinitesimal length joining the branch points at $s_\pm(\lambda^2_b)$. The physical origin of this pseudo-pole corresponds to the limit of the exchanged $m$ particle going on-shell when the $\mu$ particle becomes soft.

\section{ Phenomenology of the $Z_c(3900)$ and $Z_b(10610)$} 
In the following we apply the formalism described above to the case of the two isovector, $Z$ states, the $Z_c(3900)$ and the $Z_b(10610)$.  The $Z_c(3900)$  is observed as a peak in the $J/\psi \pi $  mass distribution near the $D\bar D^*$ threshold. The effect is seen in the decay of $Y(4260) $ to $J/\psi \pi \pi$. The $Z_b(10610)$ corresponds to a peak in $\Upsilon(nS) \pi$, $(n=1,2,3)$ near the $B\bar B^*$ threshold in the reaction $\Upsilon(5S) \to \Upsilon(1S) \pi\pi$. What follows is only an illustration and not a comprehensive study. For realistic comparison with the experimental data it would be necessary to include background contribution as has been done, for example in ~\cite{Chen:2013coa}.   

\subsubsection{ $Z_c(3900)$} 
In the notation of Sec.~\ref{sec:model} we choose (in units of GeV) $M=4.260$, $m = (m_D + m_{D^*})/2=1.936$. This gives $\lambda^2_c = 5.33$ and $\lambda^2_b=5.40$, respectively. The $t$ channel exchange of a single $D$ or $D^*$ meson corresponds to $\lambda^2 = 3.49$ and $4.03 $ respectively {\it i.e.} the single pole in the $t$-channel gives 
 a function $Q(s)$ with branch points in the region {\it iv)} and therefore is not expected to significantly enhance the amplitude near the $Z_c$ peak.

\begin{figure}
\hspace*{-.5cm}\rotatebox{270}{\scalebox{0.3}[0.3]{\includegraphics{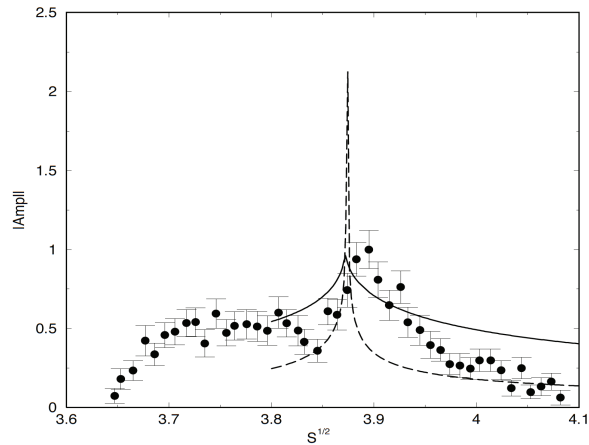}}}
\caption{ Magnitude of the amplitude $A(s)$ for $Y(4260) \to J/\psi \pi\pi$. The dashed line corresponds to the narrow width approximation of Eq.~(\ref{pole}) and the solid line corresponds to Eq.~(\ref{c1}). The amplitude and the data are normalized  
 to unity at threshold}  
\label{dd}
\end{figure}

\begin{figure}
\hspace*{-.5cm}\rotatebox{270}{\scalebox{0.3}[0.3]{\includegraphics{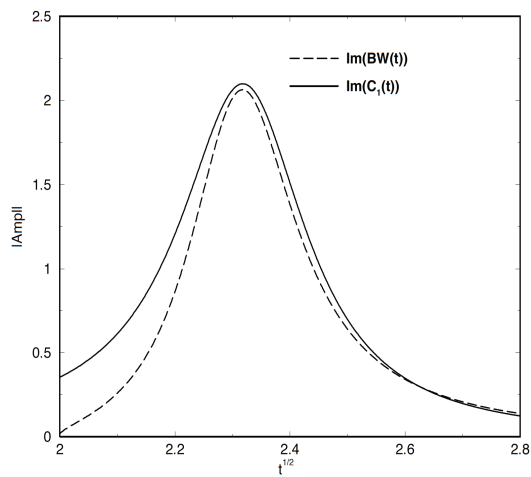}}}
\caption{Comparison between $C_1$ given by Eq.~(\ref{c1}) with the imaginary part of 
 the Breit-Wigner formula, $\propto [ m^2 - t - m \Gamma]^{-1}$ 
 with PDG values for mass $m=2.318$ and width $\Gamma=0.267$ of the $D_0^*$. Normalization of the amplitudes is arbitrary. } 
\label{dbw}
\end{figure}

At larger values of $\lambda^2$  the next $t$-channel singularity of $C$ corresponds to 
   exchange of a $D\pi$ or a $D^*\pi$ system. 
   In addition to the continuum these two-body systems  have several resonances. The lowest resonance is 
   the $D_0^*(2400)$. Ignoring the $D\pi$ continuum, {\it i.e.} using the narrow width approximation, one obtains, 
      \begin{equation} 
   C_1^{\mbox{Pole}} = \delta(m_{D^*_0}^2 - \lambda^2). \label{pole} 
   \end{equation} 
 The $D_0^*$ exchange falls entirely into the region {\it iii)}, ($\lambda^2=5.37$) and thus produces a sharp peak at $s_+(5.37) = (3.87)^2$. This is shown in Fig.~\ref{dd}. In a realistic case, however,  the $D\pi$ channel should by represented by 
       a distribution that takes into account the $D\pi$ continuum and the resonances. For example, one can approximate the  $C_1$ spectral function  by 
    \begin{equation} 
    C_1(s,\lambda^2) = \mbox{Im}\Pi(\lambda^2)  \label{c1}
    \end{equation} 
    where 
\begin{equation}     
    \Pi(\lambda^2) =  \left[ \bar m^2_{D^*_0} - \lambda^2 - \frac{1}{\pi} \int_{t_{tr}}^\infty dt' \frac{\rho(t') N(t') }{t' -\lambda^2 - i \epsilon} \right]^{-1}, 
    \end{equation} 
     $t_{tr} = (m_{D_0}+m_\pi)^2$. Choosing  $\bar m_{D^*_0} = 2.67$ and $N(t) = g \Lambda^2/(t +\Lambda^2)$ with $g=1.5$, $\Lambda=10$,  Eq.~(\ref{c1}) reproduces the  Breit-Wigner line shape with parameters corresponding to the physical values for the mass and the width of the $D^*_0$~\cite{PDG}. The comparison between $\Pi(t)$ and the Breit-Wigner formula is shown in Fig.~\ref{dbw}. 
      The reason why $C_1$ has no $s$-dependence in this model is because we ignored spin of the $t$-channel 
      $D\pi$ system. This is justifiable since $s$-dependence from the spin of an exchanged object is a smooth function. 
  Integration over the $t$-channel mass distribution of the $D\pi$ spectrum,  as expected, has the effect of smearing the sharp peak obtained in the narrow resonance approximation.   Nevertheless the cusp at threshold remains enhanced by presence of the nearby  branch points $Q(s)$. 
    There are other $D$-mesons that can contribute in the mass region of the $Z_c(3900)$ and in a phenomenological analysis of the data should be considered.  For example the $D_1(2420)$ is much more narrow that the $D_0(2400)$, however, its nominal mass places it  in region {\it ii)}.

\begin{figure}
\hspace*{-.5cm}\rotatebox{270}{\scalebox{0.3}[0.3]{\includegraphics{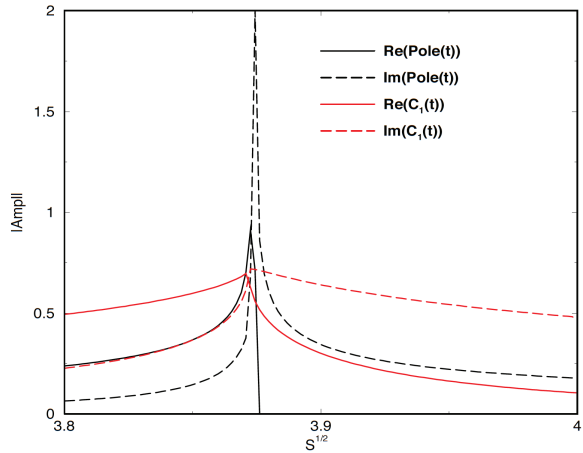}}}
\caption{Real and imaginary part of $A(s)$ computed using Eqs.~(\ref{pole}) (labeled pole) and using  Eq.~(\ref{c1}), labeled $C_1$.   } 
\label{amp}
\end{figure}

      In Fig.~\ref{amp} we compare the real and the imaginary part of $A(s)$ obtained in the narrow width approximation, ({\it cf.} Eq.~(\ref{pole})) with those computed using Eq.~(\ref{c1}). The amplitudes are different from the case when 
           the left hand cuts are far away from the threshold branch points ({\it cf.} Fig.~\ref{bb}). The numerator under the integral in Eq.~(\ref{unit}) is a complex function of  $s'$ and so is the amplitude below threshold, $s< 4m^2$. Above threshold,  in the narrow width approximation, for $\lambda^2$ in region {\it iii)}, the branch points of the numerator are located on the opposite side of the real axis producing an imaginary amplitude for  $s> \lambda^2_b$.

 \subsubsection{ $Z_b(10610)$} 
\begin{figure}
\hspace*{-.5cm}\rotatebox{270}{\scalebox{0.3}[0.3]{\includegraphics{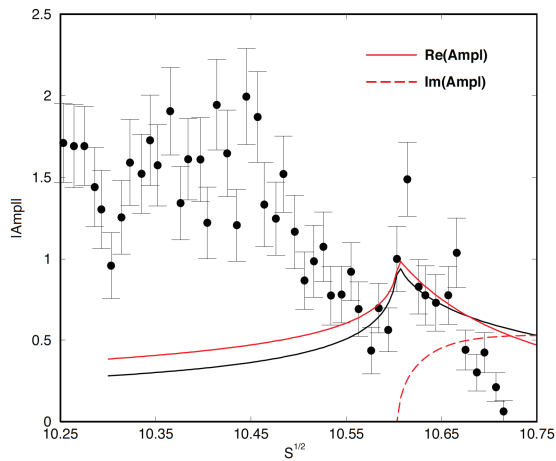}}}
\caption{Modulus (solid black), real (solid red) and imaginary (dashed red) parts of the amplitude for $\Upsilon(5S) \to \Upsilon(1S)\pi\pi$ obtained using the formula of Eq.~(\ref{c1}) with $\bar m_{D_0^*} \to \bar m_{B^*_J} = 6.036$ and $g=7.5$. As discussed in the text, the relevant range of $\lambda^2$ corresponds to the region {\it ii)} which produces a singularity in the amplitude located to the left of the unitary branch point. The real and imaginary parts thus exhibit behavior that is  typical
 for an  amplitude with a left hand side singularity.  } 
\label{bb}
\end{figure}

\begin{figure}
\hspace*{-.5cm}\rotatebox{270}{\scalebox{0.3}[0.3]{\includegraphics{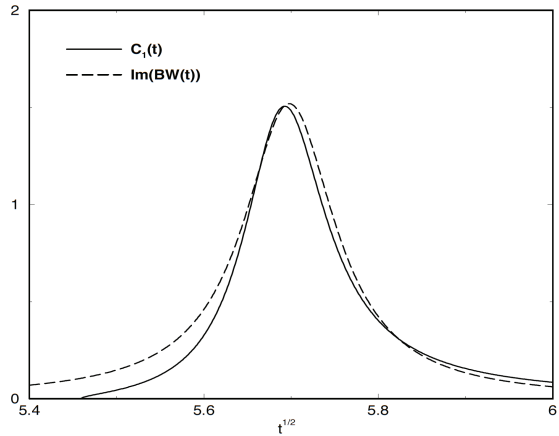}}}
\caption{Comparison of the formula in Eq.~(\ref{c1}) for the bottomonium with the imaginary part of the  Breit-Wigner formula obtained using the PDG parameters $m_{B^*_J} = 5.698$, $\Gamma_{B^*_J} =  0.128$. } 
\label{bwbb}
\end{figure}

      We apply the same analysis to the decay  $\Upsilon(5S) \to \Upsilon(1s) \pi\pi$. In this case we use $M= 10.86$, and $m=(m_B + m_{B^*})/2 = 5.302$. The $t$ channel exchange of $B$ or $B^*$ corresponds to $\lambda^2=27.87$ and $\lambda^2=28.36$, respectively, which are below $\lambda^2_c=30.86$. That is,  just as in the case of the $Z_c(3900)$,  
      one-particle-exchange contributions fall into the  region ${\it iv)}$ where threshold cusp is not enhanced. In the relevant mass range,   the $B\pi$ or $B^*\pi$ spectrum have only one known resonance, the $B^*_H$ with mass $m = 5.698$ and width  $\Gamma=0.128$. In the narrow width approximation this corresponds to $\lambda^2= 32.47$ and is above the pinch region, which corresponds to a narrow mass window between,  $30.86 <\lambda^2 < \lambda_b^2 = 30.89$.  It is only the low mass tail of the $B\pi$ distribution that overlaps with this very small region where the pinch singularity of $A(s)$ occurs. For example, the mass of the $B\pi$ system where pinching overlaps with threshold corresponds to $\lambda=\lambda_c=5.555$, which is less than one resonance width,  $\Gamma$,  below the Breit-Wigner mass of the  $B*_J$ resonance. Therefore, even though the resonance is away from the pinch region it can nevertheless enhance the normal threshold cusp. This is illustrated in Fig.~\ref{bb}. The comparison between Eq.~(\ref{c1}) with $B\pi$ system parameters and the corresponding 
        Breit-Wigner formula is shown in Fig.~\ref{bwbb}.

      \subsection{Summary} 
      Motivated by the recent discovery of the XYZ peaks in charmonium and bottomonium spectra and by theoretical  
       models that propose to explain their origin in terms of normal threshold cusp enhancements,  we preformed a systematic analysis of 2-to-2 amplitudes in the kinematic region of interest.  We found that amplitudes containing quarkonium with mass above the open flavor threshold, the amplitude $C$ in notation of Sec.~\ref{sec:model},  can have singularities in the $s$-plane that enhance the threshold cusp. For the enhancement to occur, the absorptive part of the amplitude has to have a singularity on the second Reimann sheet close to the physical region.  
     When this happens the  dispersive integral develops a pinch singularity and when the 
     singularity lying below the real axis coincides with threshold, it enhances the normal threshold cusp. 
             This, however, only occurs  
       provided the $t$-channel spectral function is large in a narrow window of masses corresponding to the region {\it iii)}.   An amplitude with the quarkonium mass below the open flavor meson ({\it i.e.} the  amplitude $B$) is not expected to be significantly enhanced. This  can be understood by considering, for example, the case $M=2m$ as discussed in Sec.~\ref{sec:model}. We thus find that in the kinematical region studied here,  
        threshold cusps are enhanced by the same type of left hand cut singularities as present in triangle graphs. Such a  mechanism has been proposed in ~\cite{Bugg:2011jr}. In Ref.~\cite{Swanson:2014tra} both the $B$ and the $C$ amplitudes were assumed to be enhanced, contrary to what is expected based on the arguments presented here. Furthermore, in the the previous studies the amplitudes were given an analytical form that is not physical. 
                In relation to the XYZ physics, triangle diagrams were used in ~\cite{Chen:2011pv,Chen:2011xk,Chen:2013coa}. 
                 These works did not, however, present a systematic study to explain which exchanges are relevant. In fact, it was assumed that it is the exchanges in the amplitude $B$ and not  $C$ that are important.  For example in \cite{Chen:2013coa} an exchange corresponding to our amplitude $B$ was found to produce an effect that could in fact  describe the data quite well. It would be interesting to analyze the singularity structure of that  amplitude and compare with our predictions. We have performed  a preliminary analysis of  $J/\pi$ and $\Upsilon(1S)\pi$ amplitudes and shown that the corresponding triangle singularities can potentially produce enhancement in the amplitude qualitatively consistent with the data. It should be noted, however, that such a mechanism cannot be considered as the  final answer. If singularities of $C$ were far away to the left from the physical region they would represent a genuine "force" {\it i.e.} virtual particle exchange, an approximation to the left hand cut that can be unitized in a standard way~\cite{mandelstam}. This point was also raised in 
                 ~\cite{FKGuo} against findings obtained with the model of ~\cite{Swanson:2014tra}. 
                                  The situation is however, even more complex since singularities in the physical region found in the triangle diagrams are due to opening of new channels and these will have to be taken into account in the unitary relation.

\section*{Acknowledgments}
I would like to thank J.~Dudek, R.L~Jaffe, M.R.~Pennington, and M.~Shepherd for useful discussions and M.~Jander and V.~Mathieu for comments on the manuscript.  This material is based upon work supported by the U.S. Department of Energy, Office of Science, Office of Nuclear Physics under contract DE-AC05-06OR23177. It is  also supported in part by the U.S. Department of Energy under Grant No. DE-FG0287ER40365.

\end{document}